\documentclass[aps,prl,twocolumn,showpacs,amsmath,amssymb,superscriptaddress]{revtex4}
\usepackage{graphicx}
\usepackage{amssymb}
\usepackage{natbib}

\begin{document}

\title{Spin excitation anisotropy as a probe of orbital ordering
in the paramagnetic tetragonal phase of superconducting BaFe$_{1.904}$Ni$_{0.096}$As$_{2}$}
\author{Huiqian Luo}
\affiliation{Beijing National Laboratory for Condensed Matter
Physics, Institute of Physics, Chinese Academy of Sciences, Beijing
100190, China}
\author{Meng Wang}
\affiliation{Beijing National Laboratory for Condensed Matter
Physics, Institute of Physics, Chinese Academy of Sciences, Beijing
100190, China}
\author{Chenglin Zhang}
\affiliation{Department of Physics and Astronomy, Rice University, Houston, Texas 77005, USA}
\affiliation{ Department of Physics and Astronomy,
The University of Tennessee, Knoxville, Tennessee 37996-1200, USA }
\author{Louis-Pierre Regnault}
\affiliation{SPSMS-MDN, UMR-E CEA/UJF-Grenoble 1, INAC, Grenoble, F-38054, France}
\author{Rui Zhang}
\affiliation{Beijing National Laboratory for Condensed Matter
Physics, Institute of Physics, Chinese Academy of Sciences, Beijing
100190, China}
\author{Shiliang Li}
\affiliation{Beijing National Laboratory for Condensed Matter
Physics, Institute of Physics, Chinese Academy of Sciences, Beijing
100190, China}
\author{Jiangping Hu}
\affiliation{Beijing National Laboratory for Condensed Matter
Physics, Institute of Physics, Chinese Academy of Sciences, Beijing
100190, China}
\affiliation{Department of Physics, Purdue University, West
Lafayette, Indiana 47907, USA}
\author{Pengcheng Dai}
\email{pdai@rice.edu}
\affiliation{Department of Physics and Astronomy, Rice University, Houston, Texas 77005, USA}
\affiliation{ Department of Physics and Astronomy,
The University of Tennessee, Knoxville, Tennessee 37996-1200, USA }
\affiliation{Beijing National Laboratory for
Condensed Matter Physics, Institute of Physics, Chinese Academy of
Sciences, Beijing 100190, China}

\date{\today}

\begin{abstract}

We use polarized neutron scattering to demonstrate that in-plane spin excitations in electron doped
superconducting BaFe$_{1.904}$Ni$_{0.096}$As$_{2}$ ($T_c=19.8$ K)
change from isotropic to anisotropic in the tetragonal phase well above
the antiferromagnetic (AF) ordering and tetragonal-to-orthorhombic  lattice distortion
temperatures ($T_N\approx T_s=33\pm 2$ K) without an uniaxial pressure.
While the anisotropic spin excitations are not sensitive to the AF order and tetragonal-to-orthorhombic lattice distortion, superconductivity induces further anisotropy for spin excitations along the
$[1,1,0]$ and $[1,-1,0]$ directions.  These results indicate
that the spin excitation anisotropy is a probe of the electronic anisotropy or orbital ordering in the
tetragonal phase of iron pnictides.
\end{abstract}

\maketitle

Understanding the electronic anisotropic state (electronic nematicity) at a temperature associated with the pseudogap phase
is one of the most important unresolved problems in the quest for mechanism of high-$T_c$ superconductivity
in copper oxides \cite{fradkin}.  For iron pnictide superconductors derived from electron-doping to
their antiferromagnetic (AF) parent compounds \cite{kamihara,cruz,dai}, there is considerable evidence
for an anisotropic electronic state in the AF phase with an orthorhombic lattice distortion
\cite{jzhao09,tmchuang,shimojima}.
Upon warming to above the AF order ($T_N$) and orthorhombic lattice distortion ($T_s$) temperatures,
iron pnictide superconductors become paramagnetic tetragonal metals \cite{dai}.
Although transport \cite{fisher}, resonant ultrasound \cite{fernandes10},
angle-resolved photoemission spectroscopy (ARPES) \cite{myi}, neutron scattering \cite{harriger},
and magnetic torque \cite{kasahara} measurements
 suggest an electronic anisotropy in the paramagnetic tetragonal phase,
much is unclear about its microscopic origin.  In one class of models, the observed electronic anisotropy
in the paramagnetic tetragonal phase of
iron pnictides \cite{fisher,fernandes10,myi,harriger,kasahara} may arise
from either in-plane spin anisotropy (spin nematic phase) \cite{jphu} as
suggested from magnetic anisotropy in torque measurements \cite{kasahara},
 or orbital ordering  \cite{cclee,kruger,lv,mdaghofer,ccchen,valenzeula} as implied from
the energy splitting of the $d_{xz}$- and $d_{yz}$-dominated bands above $T_N$ in ARPES \cite{myi}.
However, there is no sufficient experimental evidence for spin nematic phase \cite{nakajima} and
the observed orbital anisotropy in ARPES \cite{myi} may also be an extrinsic effect due to
an uniaxial pressure induced increase in $T_N$
 \cite{dhital}.  Instead of an electronic anisotropic spin nematic state or orbital ordering,
the large resistivity anisotropy seen in electron-doped BaFe$_{2-x}$Co$_x$As$_2$ \cite{fisher} has been interpreted as due to
anisotropic impurity scattering of Co-atoms in the
FeAs layer \cite{ishida,allan}.  Since the in-plane resistivity anisotropy in charge transport property  does not directly couple to spin and orbital
order, these experimental results still leave open the question concerning the presence of spin nematicity or orbital ordering in the tetragonal
phase of iron pnictides \cite{jphu,cclee,kruger,lv,mdaghofer,ccchen,valenzeula}.

Here we use polarized neutron scattering to study the spin anisotropy in electron-doped iron pnictide superconductor BaFe$_{1.904}$Ni$_{0.096}$As$_2$ ($T_c=19.8$ K) \cite{hqluo12}.  This material has incommensurate AF order ($T_N$) and tetragonal-to-orthorhombic lattice distortion ($T_s$) temperatures below
$T_N\approx T_s=33\pm 2$ K (Fig. 1) \cite{xylu}.
Since the spin anisotropy in iron pnictide must originate from a spin-orbit coupling \cite{olly},
its temperature dependence can  provide direct information on any change of electronic physics involving spin or orbital degree of freedom.
We demonstrate that spin excitations in BaFe$_{1.904}$Ni$_{0.096}$As$_2$
exhibit an in-plane isotropic to anisotropic transition
in the tetragonal phase at a temperature corresponding
to  the onset of in-plane resistivity anisotropy \cite{fisher}. While
the spin anisotropy shows no anomaly across $T_N$ and $T_s$, it enhances
dramatically below $T_c$
revealing its connection to superconductivity.
Since similar spin anisotropy is only observed in the AF orthorhombic phase of the undoped BaFe$_2$As$_2$ \cite{qureshi},
spin-orbit coupling in the paramagnetic tetragonal phase of
 BaFe$_{1.904}$Ni$_{0.096}$As$_2$ must be stabilized by an electronic anisotropic (nematic) phase or orbital ordering.

\begin{figure}[t]
\includegraphics[scale=.52]{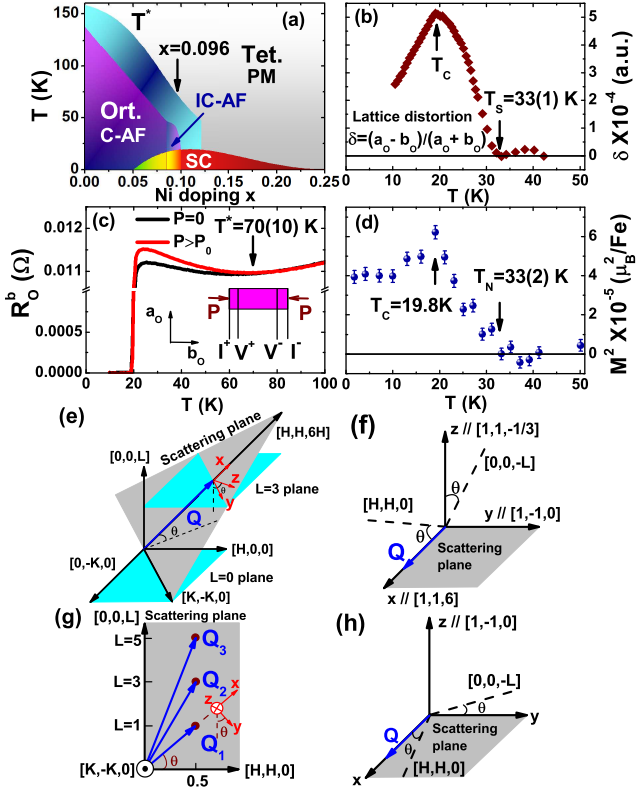}
\caption{
(a)Electronic phase diagram of
BaFe$_{2-x}$Ni$_{x}$As$_{2}$ as a function of Ni doping $x$, where $T^\ast$ is the zone boundary of anisotropic in-plane resistivity obtained from ref. \cite{Kuo}. The arrow indicates the doping level $x=$ 0.096 for our experiments.
 (b) Orthorhombic lattice distortion order parameter
 $\delta$ shows $T_s= 33 \pm 1$ K.  The
high resolution X-ray diffraction on nuclear peak $(2, 2, 12)$ experiment was from Ref. \cite{xylu}.
  (c)In-plane resistance under zero and finite uniaxial stress $P$
along $b_o$, where $P=P_0$ is the detwinned pressure. From separate neutron scattering measurements, we know
that $T_N$ and $T_s$ are uniaxial stress independent.
  (d) Temperature dependence of the AF order parameter shows $T_N = 33 \pm 2 $ K.
 (e,g) Scattering plane and neutron polarization directions in our
 experiments. (f,h) Magnetic response of SF channels in the neutron polarization analysis.}
\end{figure}

Figure 1(a) shows the schematic electronic phase diagram of BaFe$_{2-x}$Ni$_{x}$As$_2$ as determined from neutron scattering
\cite{hqluo12} and transport measurements \cite{Chen,Kuo}.   In the tetragonal phase above the $T_N$ and $T_s$, transport measurements show anisotropic resistivity  along the
orthorhombic $a_o$/$b_o$ directions below the electronic nematic ordering temperature $T^\ast$ \cite{fisher}.  We chose to study
BaFe$_{1.904}$Ni$_{0.096}$As$_{2}$ because this sample has coexisting short-range incommensurate AF order and superconductivity \cite{hqluo12}.
From previous high-resolution X-ray diffraction experiments on BaFe$_{2-x}$Co$_{x}$As$_2$ \cite{nandi} and BaFe$_{2-x}$Ni$_{x}$As$_2$
\cite{xylu}, we know that BaFe$_{1.904}$Ni$_{0.096}$As$_{2}$ changes from tetragonal to
orthorhombic lattice structure below $T_s$, and the lattice orthorhombicity becomes smaller on entering the superconducting state.  Figure 1(b) shows the temperature dependence of orthorhombicity $\delta=(a_o-b_o)/(a_o+b_o)$, revealing $T_s=33 \pm 1$ K \cite{xylu}.  Although the orthorhombicity of the system clearly decreases on cooling below $T_c$, its lattice structure does not become fully tetragonal at 10 K [Fig. 1(b)].
Similarly, temperature dependence of the magnetic order parameter indicates a N$\rm \acute{e}$el temperature
of $T_N=33 \pm 2$ K [Fig. 1(d)] \cite{hqluo12}.  To confirm the anisotropic resistivity in the tetragonal phase of
BaFe$_{1.904}$Ni$_{0.096}$As$_{2}$, we have also carried out resistivity measurements on a detwinned sample.
The outcome shows clear resistivity anisotropy for temperatures below $T^\ast=70\pm 10$ K [Fig. 1(c)].

\begin{figure}[t]
\includegraphics[scale=.5]{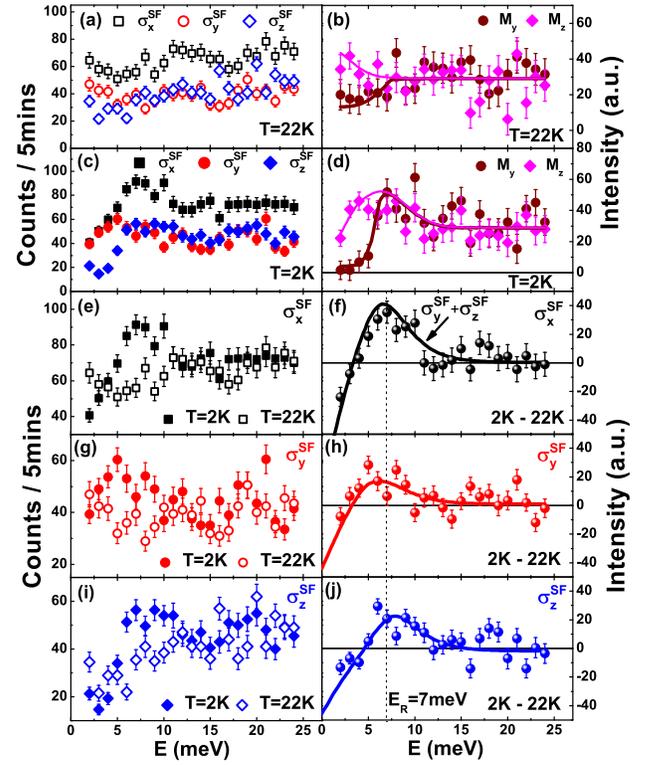}
\caption{(a) Energy scans at ${\bf Q}=(0.5, 0.5, 3)$ for SF scattering at 22 K above $T_c$ for different neutron polarization directions, marked as $\sigma_{x,y,z}^{\textrm{SF}}$.
(b) The magnetic response $M_{y}$ and $M_{z}$ extracted from (a). (c) and (d) Identical energy scans at 2 K below $T_c$ in the neutron SF channel and $M_{y}$, $M_{z}$, respectively.
(e) The total neutron SF scattering $\sigma_{x}^{\textrm{SF}}$
at 2 K and 22 K and (f) their difference, where a neutron spin resonance
is seen at $E_r=7$ meV.  (g) The $\sigma_{y}^{\textrm{SF}}$ at 2 K and 22 K and (h) their
difference. (i, j) Identical scans for $\sigma_{z}^{\textrm{SF}}$.  The solid lines in (b, d, h, j) are
guides to the eyes, and in (f) the solid line is the sum of (h) and (j).
 }
 \end{figure}

We prepared sizable high quality single crystals of BaFe$_{1.904}$Ni$_{0.096}$As$_{2}$
using self-flux method \cite{Chen} and coaligned $\sim 11$ g single crystals within $3^\circ$ full width at half maximum (FWHM). Our polarized neutron scattering experiments were
carried out using IN22 thermal triple-axis spectrometer at the Institut
Laue-Langevin, Grenoble, France \cite{olly}. The scattering planes are $[H,H,6H] \times [K,-K,0]$ and $[H,H,0]\times [0,0,L]$
 to probe the wave vector dependence of spin excitations along different directions.
Using pseudo-tetragonal lattice unit cell with $a\approx b\approx 3.956$ \AA, and $c=12.92$ \AA, the vector \textbf{Q} in three-dimensional reciprocal space in \AA$^{-1}$ is defined as $\textbf{Q}=H\textbf{a} ^*+K\textbf{b} ^*+L\textbf{c} ^*$, where $H$, $K$, and $L$ are Miller indices and
$\textbf{a} ^*=\hat{\textbf{a}}2\pi/a, \textbf{b} ^*=\hat{\textbf{b}}2\pi/b, \textbf{c} ^*=\hat{\textbf{c}}2\pi/c$
are reciprocal lattice units. We define neutron polarization directions as $x,y,z$, with $x$ parallel to \textbf{Q}, $y$ and $z$ perpendicular to \textbf{Q} as shown in Figs. 1(e)
and 1(g).  At the AF wave vector $\textbf{Q}=(0.5,0.5,3)$,
 neutron polarization directions
$x$ and $y$ are parallel to the $\textbf{Q}=[1,1,6]$ and $[1,-1,0]$ respectively, while $z$ is perpendicular to the
$[H,H,6H] \times [K,-K,0]$
scattering plane along the $\textbf{Q}=[1,1,-1/3]$ direction [Figs. 1(e) and 1(f)].  In the $[H,H,L]$ scattering plane, we probe
AF wave vectors ${\bf Q_1}=(0.5,0.5,1)$, ${\bf Q_2}=(0.5,0.5,3)$, ${\bf Q_3}=(0.5,0.5,5)$, where neutron polarization
directions $x$, $y$, and $z$ are shown in Fig. 1(g).

Since neutron scattering is only sensitive to magnetic scattering component perpendicular to the momentum transfer
\textbf{Q}, magnetic responses within the $y-z$ plane ($M_y$ and $M_z$) can be measured by using different neutron spin directions [Figs. 1(f) and 1(h)].
At a specific momentum and energy transfer, scattered neutrons can have polarizations antiparallel (neutron
spin flip or SF, $\uparrow\downarrow$) to the incident neutrons.
Therefore, the three neutron SF scattering cross sections can be
written as $\sigma_{\alpha}^{\rm SF}$, where $\alpha=x,y,z$.  The magnetic moments $M_y$ and $M_z$ can be extracted via
$\sigma_x^{\textrm{SF}}-\sigma_y^{\textrm{SF}}=cM_y$ and
$\sigma_x^{\textrm{SF}}-\sigma_z^{\textrm{SF}}=cM_z$,
where $c=(R-1)/(R+1)$ and the flipping ratio $R$ is measured by the leakage of NSF nuclear Bragg peaks into the magnetic SF channel $R=\sigma_{Bragg}^\textrm{NSF}/\sigma_{Bragg}^\textrm{SF} \approx 15$ \cite{olly}.

In previous polarized neutron scattering experiments on optimally electron-doped iron pnictide
superconductor BaFe$_{1.9}$Ni$_{0.1}$As$_2$ \cite{olly} and
BaFe$_{1.88}$Co$_{0.12}$As$_2$ \cite{steffens} without static AF order, low-energy spin excitations were found to be anisotropic
in the superconducting state.  For electron-overdoped BaFe$_{1.85}$Ni$_{0.15}$As$_2$, spin excitations are isotropic in both the normal and superconducting states \cite{msliu12}.  Figure 2(a) shows energy scans at $\textbf{Q}=(0.5,0.5,3)$ for all three SF channels ($\sigma_{\alpha}^{\rm SF}$) at $T=22$ K.  For a pure isotropic paramagnetic scattering, one expects $\sigma_{x}^{\rm SF}=2\sigma_{y}^{\rm SF}=2\sigma_{z}^{\rm SF}$ assuming
a small (negligible) background scattering \cite{olly,steffens}.
While this is indeed the case for $E\ge 5$ meV, there is apparent spin anisotropy for $E<5$ meV with $\sigma_{y}^{\rm SF}>\sigma_{x}^{\rm SF}/2>\sigma_{z}^{\rm SF}$ [Fig. 2(a)].  On cooling to $T=2$ K, the spectra are re-arranged [Fig. 2(c)].   While there is a clear resonance at $E_r\approx 7$ meV
in the $\sigma_{x}^{\rm SF}$ channel at the expense of lower energy spin excitations [Figs. 2(e)],
$\sigma_{y}^{\rm SF}$ and $\sigma_{z}^{\rm SF}$ respond to superconductivity very differently.
Instead of showing suppressed spin fluctuations below 4 meV as in the temperature difference plot for $\sigma_{x}^{\rm SF}$,
superconductivity induces a very broad resonance in $\sigma_{y}^{\rm SF}$ with magnetic intensity gain from 3 to 10 meV
[Figs. 2(g) and 2(h)].  This is similar to the $c$-axis polarized spin excitations of
BaFe$_{1.9}$Ni$_{0.1}$As$_2$ below $T_c$ \cite{olly} and BaFe$_{1.88}$Co$_{0.12}$As$_2$ \cite{steffens}.
For $\sigma_{z}^{\rm SF}$, the effect of superconductivity is to open a larger spin gap below about 5 meV
and form a resonance near $E_r=7$ meV [Figs. 2(i) and 2(j)].  Since the temperature difference plots in Figs. 2(f), 2(h),
and 2(j) should contain no background, we expect $\sigma_{x}^{\rm SF}=\sigma_{y}^{\rm SF}+\sigma_{z}^{\rm SF}$.
The solid line in Fig. 2(f) shows the sum of $\sigma_{y}^{\rm SF}$ and $\sigma_{z}^{\rm SF}$, and it is indeed statistically identical to $\sigma_{x}^{\rm SF}$.

\begin{figure}[t]
\includegraphics[scale=.35]{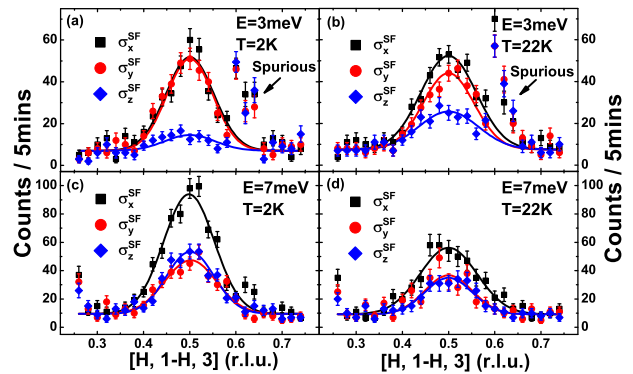}
\caption{  (a) and (b) Constant energy scans at 3 meV in the
neutron SF channel $\sigma_{x,y,z}^{\textrm{SF}}$ below and above $T_c$, respectively.
The solid lines are Gaussian fits to the data on linear backgrounds.
(c) and (d) Identical scans and results at the resonance energy of $E_r=7$ meV.
 }
\end{figure}

To quantitatively estimate the spin anisotropy from $\sigma_{\alpha}^{\rm SF}$ in Figs. 2(a) and 2(c),
we plot in Figs. 2(b) and 2(d) the energy dependence of $M_y$ and $M_z$
in the normal and superconducting states, respectively.  At $T=22$ K, the magnetic scattering show
spin anisotropy below $\sim$5 meV.  At 2 K,
the $M_y$ shows a clean spin gap below 4 meV and a resonance at $E_r=7$ meV, while $M_z$ shows a broad peak
centered around 5 meV.  In previous polarized neutron scattering experiments on electron-doped iron pnictide
superconductors \cite{olly,steffens}, similar magnetic anisotropy was found at low-energies.

Figure 3 summarizes constant energy scans along the $[H,1-H,3]$ direction at $E=3$ and 7 meV with different neutron polarizations.
At $T=2$ K, $\sigma_{x}^{\rm SF}$ and $\sigma_{y}^{\rm SF}$ at $E=3$ meV display well-defined peaks at $(0.5,0.5,3)$ with almost the
same magnitude, while $\sigma_{z}^{\rm SF}$ has only a broad weak peak center at at $(0.5,0.5,3)$  [Fig. 3(a)].  These data are
consistent with constant-$Q$ scans in Fig. 2.  At $T=22$ K, similar scans show three separate peaks satisfying
$\sigma_{y}^{\rm SF}>\sigma_{x}^{\rm SF}/2>\sigma_{z}^{\rm SF}$, again confirming the anisotropic nature of the normal state spin excitations
in Fig. 2(a).  For comparison, spin excitations at the resonance energy of $E_r=7$ meV are
completely isotropic below [Fig. 3(c)]
and above [Fig. 3(d)] $T_c$ satisfying $\sigma_{x}^{\rm SF}=2\sigma_{y}^{\rm SF}=2\sigma_{z}^{\rm SF}$.

\begin{figure}[t]
\includegraphics[scale=.4]{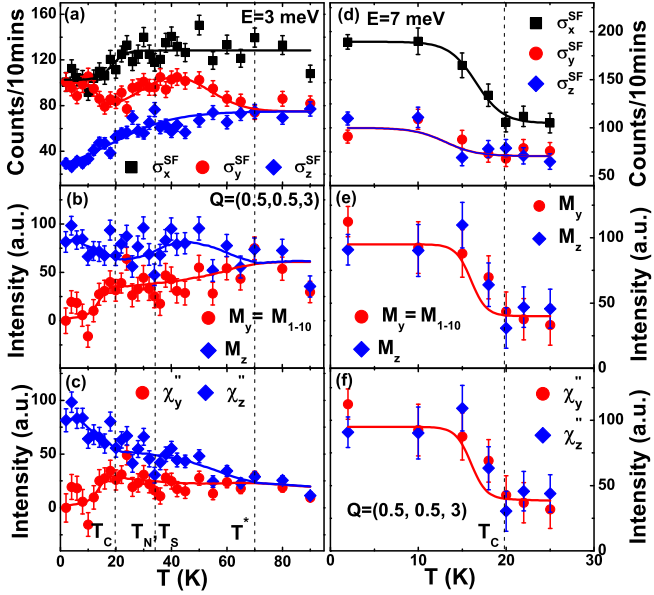}
\caption{(a) Temperature dependence of neutron SF scattering cross section $\sigma_{x,y,z}^{\textrm{SF}}$
at 3 meV and ${\bf Q}=(0.5,0.5,3)$. (b) The temperature dependence of magnetic response
along the $[1,-1,0]$ ($M_{y}$) and $[1,1,-1/3]$ ($M_{z}$) directions.
Clear anisotropy persists up to $T^\ast=$ 70 K. (c) Temperature dependence of the dynamic susceptibility,
$\chi^{\prime\prime}_y$ and $\chi^{\prime\prime}_z$.
(d),(e), and (f) Identical results at the resonance energy of $E_r=7$ meV.
 }
\end{figure}

Given the clear experimental evidence for anisotropic spin excitations at $E=3$ meV and its possible coupling to superconductivity as illustrated in Figs. 2 and 3, it
would be interesting to measure the temperature dependence of the spin anisotropy.  Figure 4(a) shows the temperature dependent scattering
for $\sigma_{\alpha}^{\rm SF}$ at ${\bf Q}=(0.5,0.5,3)$ and $E=3$ meV.
At temperatures above 70 K, we see $\sigma_{x}^{\rm SF}\approx 2\sigma_{y}^{\rm SF}\approx 2\sigma_{z}^{\rm SF}$ indicating that
spin excitations are isotropic with $M_y=M_z$.  On cooling to below 70 K, we see a clear splitting of the temperature dependent
$\sigma_{y}^{\rm SF}$ and $\sigma_{z}^{\rm SF}$.  While $\sigma_{x}^{\rm SF}$ shows no visible changes cross 70 K,
$\sigma_{y}^{\rm SF}$ increases and $\sigma_{z}^{\rm SF}$ decreases with decreasing temperature below 70 K before saturating
around 40 K.  On cooling further to crossing $T_N$ and $T_s$, there are no statistically significant changes in
$\sigma_{x}^{\rm SF}$, $\sigma_{y}^{\rm SF}$, or $\sigma_{z}^{\rm SF}$, indicating that spin anisotropy at $E=3$ meV does not
respond to AF ordering and tetragonal-orthorhombic lattice distortion.  Finally, on cooling below $T_c$,
we see a clear reduction in $\sigma_{x}^{\rm SF}$, revealing a suppression of the spin excitations for energies below the
resonance. On the other hand, while $\sigma_{y}^{\rm SF}$ increases at $T_c$ and
 merges with $\sigma_{x}^{\rm SF}$ below around 10 K,
 $\sigma_{z}^{\rm SF}$ exhibits a further reduction in intensity below $T_c$.
Figure 4(b) shows the temperature dependence of the magnetic scattering
$M_y$ and $M_z$ obtained from $\sigma_{\alpha}^{\rm SF}$.
On cooling, spin excitations first
change from isotropic to anisotropic below
approximately 70 K, and further enhance anisotropy below $T_c$ with almost zero $M_y$ at 2 K.  Figure 4(c) shows
temperature dependence of
the imaginary part of the dynamic susceptibility, $\chi^{\prime\prime}$, along the $y$ and $z$ directions.
They show again the appearance of spin anisotropy below 70 K with
no changes across $T_N$ and $T_s$, and a further spin anisotropy change below $T_c$.

\begin{figure}[t]
\includegraphics[scale=.3]{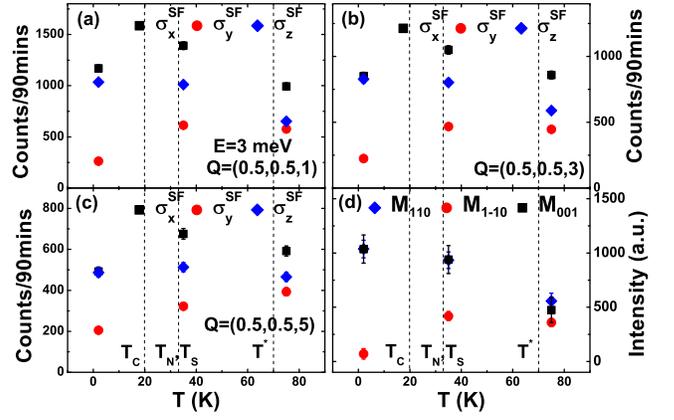}
\caption{(a,b,c) Temperature dependence of neutron SF scattering cross section $\sigma_{x,y,z}^{\textrm{SF}}$
at 3 meV and ${\bf Q_{1,2,3}}=(0.5,0.5,L)$ with $L=1,3,5$. (d) The temperature dependence of magnetic response
along the $[1,1,0]$ ($M_{110}$), $[1,-1,0]$ ($M_{1\bar{1}0}$), and $[0,0,1]$ ($M_{001}$)
 directions.  Clear in-plane anisotropy persists up to $T^\ast=$ 70 K.
 }
\end{figure}

Figure 4(d) shows temperature dependence of the magnetic intensity at the resonance energy $E_r=7$ meV.
At all measured temperatures, we find $\sigma_{x}^{\rm SF}\approx 2\sigma_{y}^{\rm SF}\approx 2\sigma_{z}^{\rm SF}$, thus confirming
the isotropic nature of the mode.  Figures 4(e) and 4(f) are the corresponding temperature dependence of $M_y$,$M_z$ and
$\chi_y^{\prime\prime}$,$\chi_z^{\prime\prime}$, respectively.  In both cases, there is intensity
increase below $T_c$, consistent
with earlier work on the resonance \cite{olly,steffens}. For comparison, we note that spin excitations in superconducting iron chalcogenides have
slightly anisotropic resonance with isotropic spin excitations below it \cite{boothroyd,prokes}.

In previous polarized neutron measurements on the parent compound BaFe$_2$As$_2$ \cite{qureshi}, it was found that the in-plane polarized spin waves exhibit a larger gap than the out-of-plane polarized ones, suggesting that it costs more energy to rotate a spin within the orthorhombic
$a$-$b$ plane than to rotate it perpendicular to the FeAs layers.  However,
the spin anisotropy immediately disappears in the paramagnetic tetragonal state above
$T_N$ and $T_s$ \cite{qureshi}.  Since $M_y$ is the spin moment in the FeAs layers [Fig. 1(e)],
the $M_y$ and $M_z$ anisotropy should also represent the spin
anisotropy along the $[1,-1,0]$ and $[1,1,-1/3]$ directions, respectively.
To determine the precise anisotropic direction of spin excitations at
$E=3$ meV, we measured
$\sigma_{\alpha}^{\rm SF}$ at ${\bf Q_{1,2,3}}$ in the $[H,H,L]$ zone [Figs. 5(a)-5(c)].
At $T=2$ K ($\ll T_c$), we see $\sigma_{x}^{\rm SF}\approx \sigma_{z}^{\rm SF}
\gg \sigma_{y}^{\rm SF}$ at all wave vectors probed.  On warming to 35 K ($>T_N, T_s$), we have
$\sigma_{x}^{\rm SF}> \sigma_{z}^{\rm SF}
> \sigma_{y}^{\rm SF}$.  At 75 K, we find
$\sigma_{x}^{\rm SF}\approx 2\sigma_{y}^{\rm SF}\approx 2\sigma_{z}^{\rm SF}$, suggesting weak or no spin anisotropy.  By considering
wave vector dependence of spin excitations in Figs. 5(a)-5(c), we estimate the temperature dependence of
$M_{110}$, $M_{1\bar{1}0}$, and $M_{001}$ [Fig. 5(d)] \cite{supplementary}.

In the superconducting
orthorhombic state, there are clear in-plane magnetic anisotropy with $M_{001}\sim M_{110} \gg M_{1\bar{1}0}\approx 0$.
In the paramagnetic tetragonal state just above $T_s$ and $T_N$, we still have strong in-plane magnetic anisotropy with
$M_{110}\sim M_{001} > M_{1\bar{1}0}$.  This is surprising because domains associated with
the in-plane AF wave vector $\textbf{Q}=(0.5,0.5)$ are randomly mixed with those associated with the $\textbf{Q}=(0.5,-0.5)$
in the tetragonal phase.  In the AF orthorhombic
state, the low-energy spin excitations associated with the $\textbf{Q}=(0.5,0.5)$ domains are well separated from those
associated with $\textbf{Q}=(0.5,-0.5)$ in reciprocal space \cite{harriger}.  If there are strong
paramagnetic scattering at $\textbf{Q}=(0.5,-0.5)$ arising
from domains associated with $\textbf{Q}=(0.5,0.5)$
in the tetragonal phase, one should not be able to determine the spin excitation anisotropy in
neutron polarization analysis.
However, recent unpolarized neutron experiments on nearly 100\% mechanically detwinned  BaFe$_{2-x}$Ni$_x$As$_2$ reveal that
spin excitations in the paramagnetic tetragonal state are still centered mostly at $\textbf{Q}=(0.5,0.5)$ \cite{xylu13}.  Therefore,
our neutron polarization analysis provides the most compelling evidence for the in-plane spin anisotropy in the paramagnetic tetragonal phase of
BaFe$_{1.904}$Ni$_{0.096}$As$_2$ [Fig. 5(d)].  Since such spin excitation anisotropy occurs at the AF wave vector
$\textbf{Q}=(0.5,0.5)$, it does not break the $C_4$ rotational symmetry of the underlying lattice.

In summary, we have discovered that an in-plane
isotropic to anisotropic spin fluctuation transition occurs in the tetragonal phase of
superconducting BaFe$_{1.904}$Ni$_{0.096}$As$_{2}$ without an uniaxial pressure, consistent with resistivity anisotropy.  The spin anisotropy is further enhanced upon entering into the superconducting state. Therefore, our experimental results establish the in-plane spin anisotropy as a new experimental probe to study
the spontaneously broken electronic symmetries in strain free iron pnictides.

 The work at IOP, CAS, is
supported by MOST (973 project: 2012CB821400, 2011CBA00110, and
2010CB833102) and NSFC (No.11004233). The work at UTK and Rice is supported by the
U.S. NSF-DMR-1063866.

\clearpage
\appendix
\section{Supplementary Materials}

\begin{figure*}
\renewcommand\thefigure{S1}
\begin{center}
\includegraphics[width=0.9\linewidth]{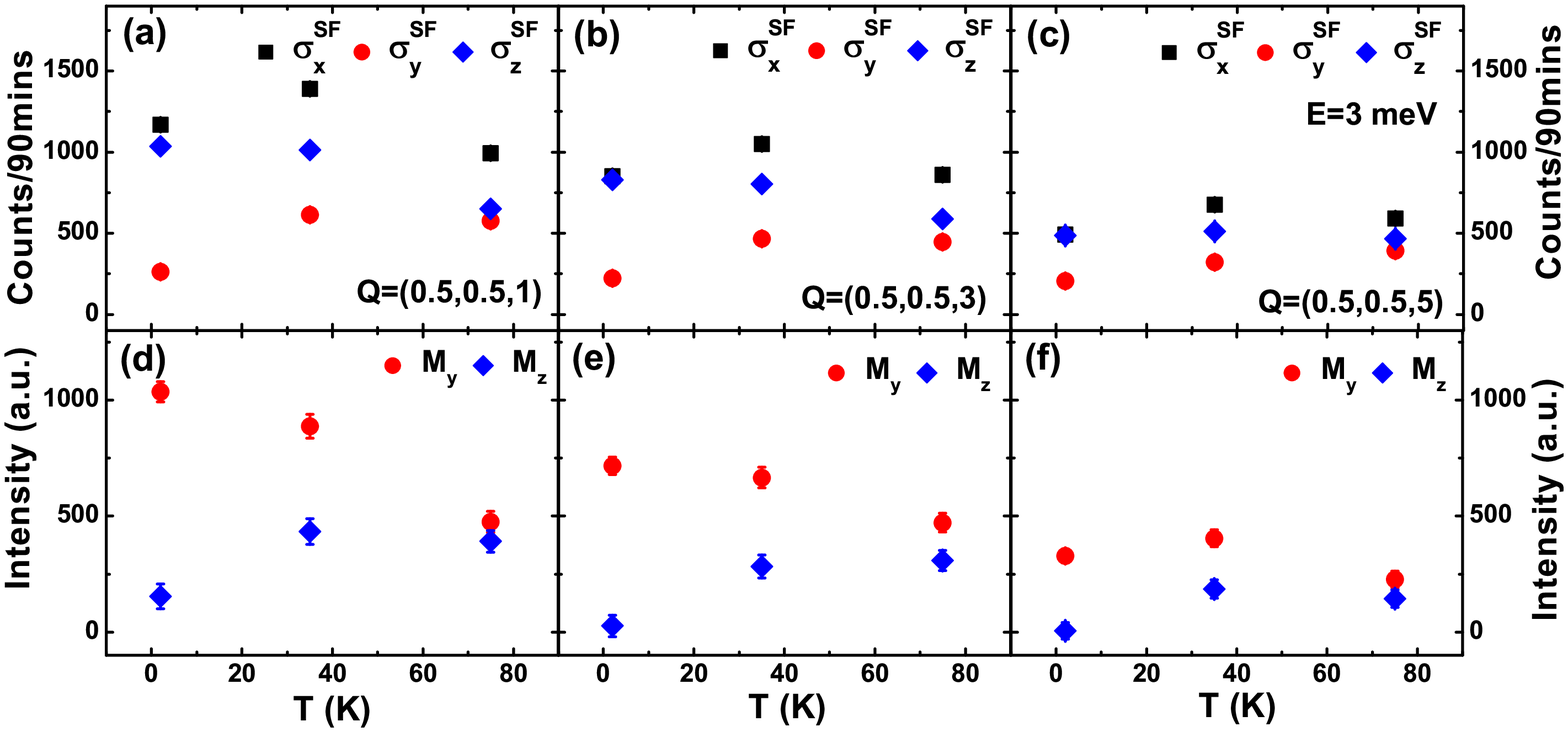}
\end{center}\caption{ (a)-(c) Temperature dependence of neutron SF scattering
cross section $\sigma_{x,y,z}^{\rm SF}$ at $E=3$ meV for ${\bf
Q_{1,2,3}}=(0.5, 0.5, 1), (0.5, 0.5, 3), (0.5, 0.5, 5)$ in the
scattering geometry shown in Fig. 1(g). (d)-(f) The magnetic
response $M_{y}$ and $M_{z}$ at the three wave vectors.
  } \label{Fig:figS1}
\end{figure*}

Figure S1 shows the raw data of neutron SF scattering cross section
$\sigma_{x,y,z}^{\rm SF}$  and the magnetic responses $M_{y}$ and
$M_{z}$ at $E=3$ meV for three measured wave vectors ${\bf
Q_{1,2,3}}=$(0.5, 0.5, 1), (0.5, 0.5, 3), (0.5, 0.5, 5),
respectively. To deduce the three components of the magnetic
excitations $M_{110}$, $M_{1\bar{1}0}$ and $M_{001}$, which
correspond to spin excitations along the $a_o$, $b_o$, and $c$-axes
of the orthorhombic lattice unit cell, we need to consider the angle
dependence of spin excitations with respect to the neutron
polarization directions [Fig. 1(g)]. For each measured wave vector
${\bf Q_{1,2,3}}$, we can estimate $M_y$ and $M_z$ from the raw data
via

\begin{equation}
\left\{
\begin{array}{ccc}
\sigma_{x}^{\rm SF}-\sigma_{y}^{\rm SF}=cM_y,\\
\\[1pt]
\sigma_{x}^{\rm SF}-\sigma_{z}^{\rm SF}=cM_z,\\
\end{array}
\right.
\end{equation}
where $c=(R-1)/(R+1)$ and the spin flipping ratio $R \approx 15$.
The components of the magnetic responses along each of the
crystallographic high symmetry directions can be written as:
$M_y=M_{110}\sin^2\theta+M_{001}\cos^2\theta$ and
$M_z=M_{1\bar{1}0}$ for each of the wave vectors, where $\theta$ is
the angle between the neutron polarization direction $x$ and the
$[1,1,0]$ direction as shown in Fig. 1(g). These results are shown
in Figs. S1(d) -(f) for wave vector ${\bf Q_{1,2,3}}$.

In our experiments, the lattice parameters are $a \approx b \approx
3.956$ \AA, and $c = 12.92$ \AA\ using the pseudo-tetragonal
structure, thus the angle $\theta$ between the wave vector ${\bf
Q}=[0.5, 0.5, L]$ and the $[H, H, 0]$ direction can be calculated by
using $\tan\theta=(2\pi L/c)/(2\pi
\sqrt{(1/2a)^2+(1/2b)^2})=\sqrt{2}aL/c$, giving the results:
$\theta_1= 23.4^{\circ}$, $\theta_2= 52.4^{\circ}$, $\theta_3=
65.2^{\circ}$ for ${\bf Q_{1,2,3}}=$(0.5, 0.5, 1), (0.5, 0.5, 3),
(0.5, 0.5, 5), respectively. Hence we have a series equations for
$M_y$ from the three probed wave vectors:
\begin{widetext}
\begin{equation}
\left\{
\begin{array}{ccc}
M_y({\bf Q_1})=M_{110}\sin^223.4^{\circ}+M_{001}\cos^223.4^{\circ}=0.16M_{110}+0.84M_{001},\\
\\[1pt]
M_y({\bf Q_2})=M_{110}\sin^252.4^{\circ}+M_{001}\cos^252.4^{\circ}=0.63M_{110}+0.37M_{001},\\
\\[1pt]
M_y({\bf Q_3})=M_{110}\sin^265.2^{\circ}+M_{001}\cos^265.2^{\circ}=0.82M_{110}+0.18M_{001}.\\
\end{array}
\right.
\end{equation}
\end{widetext}

 Since $M_{110}$ and $M_{001}$ should be the same at
these wave vectors except for the differences in the magnetic form
factor and instrumental resolution, one can in principle
unambiguously solve $M_{110}$ and $M_{001}$ if measurements at two
equivalent wave vectors are carried out.  As we can see, $M_y$
measurements at low wave vector ${\bf Q_1}$ will be more sensitive
to the $c$-axis polarized spin excitations $M_{001}$, while
identical measurements at ${\bf Q_3}$ will be more sensitive to
$M_{110}$.  Assuming that spin excitations in the system follow the
Fe$^{2+}$ magnetic form factor, we would expect that $F^2({\bf
Q_1})=0.826$, $F^2({\bf Q_2})=0.652$, and $F^2({\bf Q_3})=0.418$ for
${\bf Q_{1,2,3}}=$(0.5, 0.5, 1), (0.5, 0.5, 3), (0.5, 0.5, 5),
respectively. To estimate the contributions of instrumental
resolution at different wave vectors, we note that instrumental
contributions for spin excitations should be independent of neutron
spin polarizations.  If we assume that spin excitations
$M{1\bar{1}0}$ are identical for different wave vectors except for
the magnetic form factor and instrumental resolution, we should have
$M_{1\bar{1}0}=M_z({\bf Q_1})/[F^2({\bf Q_1}) R_1]=M_z({\bf
Q_2})/[F^2({\bf Q_2}) R_2]=M_z({\bf Q_3})/[F^2({\bf Q_3}) R_3]$,
where $R_1$, $R_2$, and $R_3$ are the scale factors representing
contributions from instrumental resolutions at these wave vectors.
Similarly, we have

\begin{equation}
\left\{
\begin{array}{ccc}
M_y({\bf Q_1})/[F^2({\bf Q_1}) R_1]=0.16M_{110}+0.84M_{001},\\
\\[1pt]
M_y({\bf Q_2})/[F^2({\bf Q_2}) R_2]=0.63M_{110}+0.37M_{001},\\
\\[1pt]
M_y({\bf Q_3})/[F^2({\bf Q_3}) R_3]=0.82M_{110}+0.18M_{001},\\
\end{array}
\right.
\end{equation}

Since we have measured SF scattering at three different wave
vectors, we have over-determined the values of $M_y$ and $M_z$.  Of
the combined scale factors $F^2({\bf Q_1}) R_1$, $F^2({\bf Q_2})
R_2$, and $F^2({\bf Q_3}) R_3$, only two are independent as we do
not measure $M_{1\bar{1}0}$ and $M_{110}$ in absolute units.
Therefore, we can accurately determine $F^2({\bf Q_2}) R_2$ and
$F^2({\bf Q_3}) R_3$ (assuming $F^2({\bf Q_1}) R_1=1$) using
measured values of $M_y$ and $M_z$.  We do not need to know the
values of the magnetic form factor. This procedure will also allow
us to unambiguously determine the temperature dependence of
$M_{110}$, $M_{1\bar{1}0}$, and $M_{001}$ as shown in Fig. 5(d). In
any case, Figs. S1(e) and S1(f) show clear differences between $M_y$
and $M_z$ at 35 K.  Since at $M_y({\bf Q_3})$ has 82\% contribution
from $M_{110}$ and $M_z$ is 100\% $M_{1\bar{1}0}$, in-plane spin
excitations are unambiguously anisotropic in the paramagnetic
tetragonal state of BaFe$_{1.904}$Ni$_{0.096}$As$_2$ without an
uniaxial pressure.

\end{document}